# Phonon heat conduction in Al$_{1-x}$Sc$_x$N thin films

Chao Yuan[1], Mingyo Park[2], Yue Zheng[2], Jingjing Shi[1], Rytis Dargis[3], Samuel Graham[1]*, Azadeh Ansari[2]*

1 George W. Woodruff School of Mechanical Engineering, Georgia Institute of Technology, Atlanta, Georgia 30332, USA
2 School of Electrical and Computer Engineering, Georgia Institute of Technology, Atlanta, Georgia 30332, USA
3 IQE NC, 494 Gallimore Dairy Rd., Greensboro, NC, 27407 USA

*Author to whom correspondence should be addressed: sgraham@gatech.edu; azadeh.ansari@ece.gatech.edu

**Abstract**
Aluminum scandium nitride alloy (Al$_{1-x}$Sc$_x$N) is regarded as a promising material for high-performance acoustic devices used in wireless communication systems. Phonon scattering and heat conduction processes govern the energy dissipation in acoustic resonators, ultimately determining their performance quality. This work reports, for the first time, on phonon scattering processes and thermal conductivity in Al$_{1-x}$Sc$_x$N alloys with the Sc content (x) up to 0.26. The thermal conductivity measured presents a descending trend with increasing x. Temperature-dependent measurements show an increase in thermal conductivity as the temperature increases at temperatures below 200K, followed by a plateau at higher temperatures (T> 200K). Application of a virtual crystal phonon conduction model allows us to elucidate the effects of boundary and alloy scattering on the observed thermal conductivity behaviors. We further demonstrate that the alloy scattering is caused mainly by strain-field difference, and less by the atomic mass difference between ScN and AlN, which is in contrast to the well-studied Al$_{1-x}$Ga$_x$N and Si$_x$Ge$_{1-x}$ alloy systems where atomic mass difference dominates the alloy scattering. This work studies and provides the quantitative knowledge for phonon scattering and the thermal conductivity in Al$_{1-x}$Sc$_x$N, paving the way for future investigation of materials and design of acoustic devices.

Aluminum scandium nitride alloy (Al$_{1-x}$Sc$_x$N) has been regarded as a great material candidate for next generation wide-bandwidth and high-frequency wireless communication devices (*e.g.* 5G filters) due to its enhanced piezoelectric properties [1] that translates into high electromechanical coupling coefficients ($k_t^2$) in acoustic resonators [2-4]. The quality factor (*Q*) of acoustic resonators is governed by the dissipation of energy stored in the resonant vibrational mode to other acoustic modes or the environment [5]. In particular, phonon transport processes in the resonator material influence the dissipation mechanisms. For instance, time-varying strain gradients drive irreversible spatial phonon transport (heat flow), known as thermoelastic dissipation (TED) [5-7]. TED is dominant in lower frequency devices and depends inversely on thermal coefficient of expansion and thermal conductivity [7]. In addition to TED, there are intrinsic loss mechanisms due to the scattering of the acoustic phonons of a resonant mode with thermal phonons. These scattering processes have been divided into two regimes, Akhiezer and Landau-Rumer damping, based upon the relative relationship between the wavelength of the propagating acoustic wave ($\lambda$) and the phonon mean free path ($l$) [5-9]. Therefore, understanding the phonon scatterings and thermal conductivities of $Al_{1-x}Sc_xN$ material will benefit for the prediction of the energy loss and the resulting quality factor (*Q*), which will ultimately help the choice of material and device design. Besides influencing the resonator *Q*, phonon conduction governs the heat removal in materials. Self-heating in resonators due to low thermal conductivity could result in an abrupt increase of the local temperature, which will degrade device performance and reliability [10, 11]. Such self-heating becomes critical for acoustic resonators under high power loads used for 5G applications [12-14].

Despite the importance, studies of phonon scattering mechanisms and thermal conductivity are completely unavailable for $Al_{1-x}Sc_xN$. Thermal conductivity of materials could routinely be obtained with the robust characterization techniques, such as thermoreflectance [15, 16] and 3-Omega methods [17-19]. However, phonon scattering, particularly for the alloy systems, have been investigated less due to the challenges associated with their quantitative determination and theoretical modelling. While the inelastic x-ray scattering methods provide the direct observation of phonon transport properties (*e.g.,* phonon dispersion, linewidths) [20, 21], they are not able to differentiate various intrinsic and extrinsic scattering mechanisms that normally exist in the complex alloy systems. In addition, phonon scattering and thermal conductivity should be investigated concurrently as they are strongly correlated. Another approach to studying phonon scattering and thermal conductivity is the theoretical method, such as the first-principles methods [22-26] and the analytical Debye-Callaway methods [27-35]. Regardless of the applied theoretical method, it is normally combined with a virtual crystal (VC) assumption that the disordered crystal is treated to be an ordered one of the average lattice parameters and phonon-related properties taken according to the composition [24-26, 31-33]. The VC model has been able to reproduce the experimental thermal conductivities of some piezoelectric alloy materials- $Si_xGe_{1-x}$ and $Al_{1-x}Ga_xN$, in which the components have the similar crystal structure. Applying the same approach to transitional Sc metal alloyed $Al_{1-x}Sc_xN$ system remain questions as the ScN rocksalt structure is distinctively different from AlN wurtzite structure.

This work reports on the experimental and theoretical investigations of the phonon transport in $Al_{1-x}Sc_xN$ alloys with x varied from 0.028 to 0.26. The thermal conductivities were measured by time-domain thermoreflectance (TDTR) [36, 37]. The analytical Debye-Callaway model combined with the virtual crystal assumption was found to match and describe the experimental thermal conductivity trends well. The analytical model was then used to analyze the phonon scattering mechanisms.

A series of $Al_{1-x}Sc_xN$ thin films, with Sc concentration from 0.028 to 0.26, were grown on Si <111> wafer using plasma-assisted molecular beam epitaxy (PA-MBE). The details of growth method can be seen in [38]. The thicknesses of $Al_{1-x}Sc_xN$ layers are 400 nm or 800 nm. The Sc concentration of the nitride layers was confirmed by X-ray photoelectron spectroscopy (XPS), with the experiment details seen in the **supplementary material**. Thin film structures were characterized with the transmission electron microscopy (TEM) and scanning transmission electron microscopy (STEM). Figure 1(a) and 1(b) show the cross-sectional TEM and STEM images of an $Al_{1-x}Sc_xN$/Si sample. Successive columnar grains, oriented normally to the $Al_{1-x}Sc_xN$/Si interface are observed. The in-plane grain size is approximately 100 nm. Columnar morphology is also seen in the sputtered $Al_{1-x}Sc_xN$ films [39, 40] and similar to the structure of polycrystalline thin-film diamond grown by MOCVD method [41, 42]. Other defects near the $Al_{1-x}Sc_xN$ /Si interface such as smaller grains, minimal twists and orientation disturbance, were characterized with the high-resolution TEM (inset of Figure 1(a)). In general, these characterizations are necessary to minimize the uncertainty in the analysis of thermal characterization data, and theoretical analysis in the phonon-transport model.

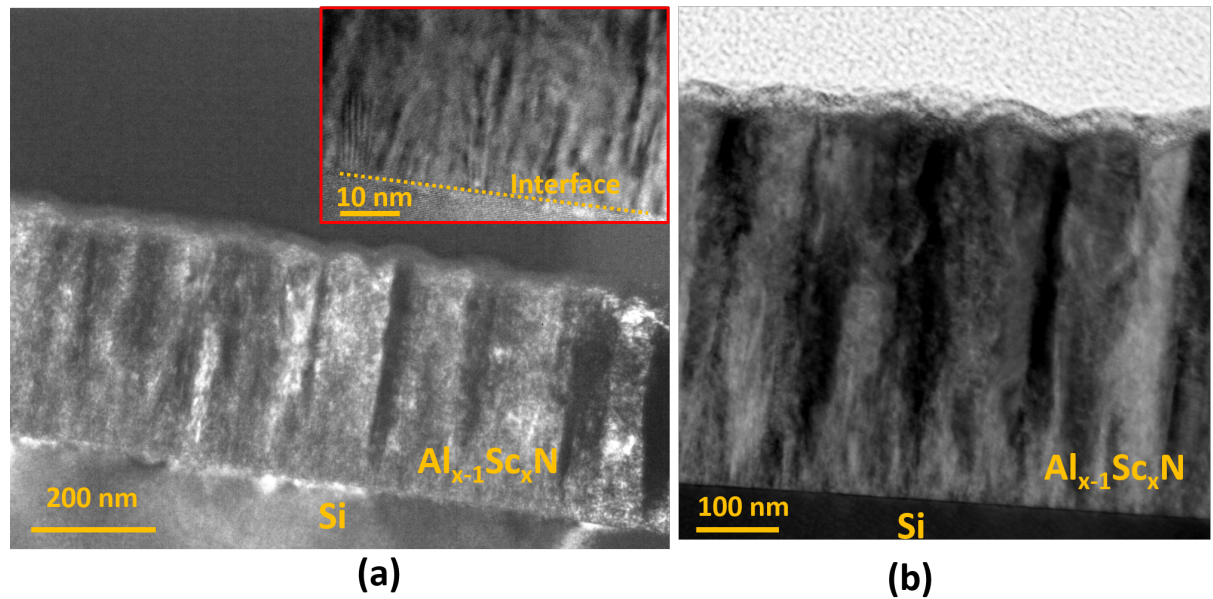

Figure 1. Cross-sectional (a) transmission electron microscopy (TEM) and (b) scanning transmission electron microscopy (STEM) images of an $Al_{1-x}Sc_xN$/Si sample under study. The inset in (a) is the high-resolution TEM imaged at the $Al_{1-x}Sc_xN$/Si interface area.

Using TDTR, we measured the room-temperature thermal conductivity for all fabricated $Al_{1-x}Sc_xN$ layers with the results shown in Figure 2(a). The details of the TDTR system and the measurement method can be found in the **supplementary material**. The thermal conductivity of $Al_{1-x}Sc_xN$ measured at room temperature ranges from 3.6 W/mK to 10.9 W/mK, approximately 10 times lower than that of crystalline AlN with similar thickness [43, 44] and 30-90 times lower than that of bulk single crystal AlN [45]. We observed a clear descent trend in thermal conductivity as x increases up to 0.26. Note that the thermal conductivity of alloy systems (such as $Si_xGe_{1-x}$ [31, 33], $Al_{1-x}Ga_xN$ [25, 32], $Mg_2Si_xSn_{1-x}$ [26]) generally follows a bathtub trend (decaying with increasing x from 0, before reaching the valley, and then increasing until x reaches 1). We did not observe the full bathtub curve behavior in the studied $Al_{1-x}Sc_xN$ samples due to the limited range over which x varied. To study whether $Al_{1-x}Sc_xN$ follows the bathtub trend, further testing of the samples with higher x is necessary. It should also be noted that with x increasing, the $Al_{1-x}Sc_xN$ structure will change from the wurtzite phase of AlN into the rocksalt phase of ScN [46]. Motivated by industrial interest in acoustic devices based on the single-phase textured wurtzite materials, this study, as well as most of other $Al_{1-x}Sc_xN$ studies, has focused on the wurtzite $Al_{1-x}Sc_xN$ with lower x values. Interestingly, the two thicker (800 nm) $Al_{1-x}Sc_xN$ layers' thermal conductivities are located within the 400 nm layers' thermal conductivity decreasing trend, as seen in Figure 2(a). This indicates the thermal conductivities of the thinned layers have not been affected by the thickness reduction. Considering the columnar morphology observed in Al$_{1-x}$Sc$_x$N, the thermal conductivities are likely limited by grain boundaries. Temperature-dependent (80K-450K) thermal conductivities were also measured for three samples (x=0.028, 0.076 and 0.173), and the results are plotted in Figure 2(b). The results show an increase in Al$_{1-x}$Sc$_x$N thermal conductivities as the temperature increases to 200K, followed by a plateau at higher temperature (T>200 K). This temperature behavior is dramatically different from pure bulk AlN, further implying that alloying and grain boundaries affect the thermal conductivities.

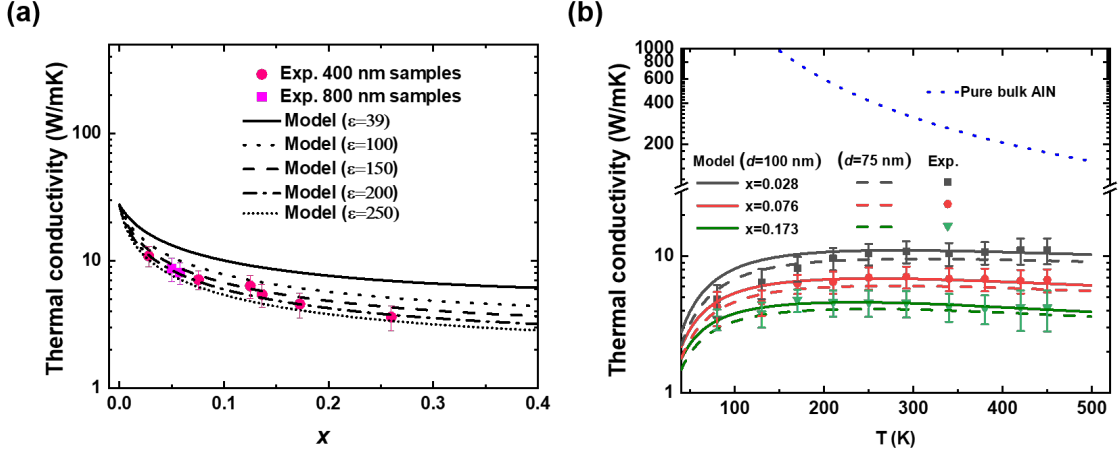

Figure 2. (a) Experimental thermal conductivities of Al$_{1-x}$Sc$_x$N with x from 0.028 to 0.26 and thickness of 400 nm and 800 nm, and comparison with a series of modelling results calculated with the phenomenological parameter ($\varepsilon$) in Eq. (5) adjusted from 39 to 250. (b) Temperature-dependent thermal conductivity measurements and models of Al$_{1-x}$Sc$_x$N with x being 0.028, 0.076 and 0.173; The model results were plotted for two different effective boundary lengths-$d$: 100 nm and 75 nm. In (b), modelling results of pure bulk AlN are also plotted for comparison.

To analyze the phonon scattering and their effects on thermal conductivities, we turn to an analytical model. Here, the thermal conductivity of the Al$_{1-x}$Sc$_x$N ($k$) was analyzed in the framework of a modified Callaway-Holland model [27-30]:

$$k = \frac{1}{3}\frac{K_B^4 T^3}{2\pi^2 \hbar^3}\sum_i v_i \left(I_{1,i} + \beta_i I_{2,i}\right) \quad (1)$$

where $K_B$ Boltzmann constant, $\hbar$ reduced Planck constant, $T$ the temperature, $i$ represents the different branches of the phonons (one longitudinal and two transverse branches, i. e. $L$, $T_1$, $T_2$, were taken into account), $v_i$ the sound velocity of the $i$-th phonon branch, and $I_{1,i}$, $I_{2,i}$ and $\beta_i$ are the following integrals:

$$I_{1,i} = \int_{x=0}^{\theta_i/T} \tau_{C,i} \frac{x^4 e^x}{(e^x-1)^2} dx \quad (2)$$

$$I_{2,i} = \int_{x=0}^{\theta_i/T} \frac{\tau_{C,i}}{\tau_{N,i}} \frac{x^4 e^x}{(e^x-1)^2} dx \quad (3)$$

$$\beta_i = \frac{\int_{x=0}^{\theta_i/T} \frac{\tau_{C,i}}{\tau_{N,i}} \frac{x^4 e^x}{(e^x-1)^2} dx}{\int_{x=0}^{\theta_i/T} \frac{\tau_{C,i}}{\tau_{N,i}\tau_{R,i}} \frac{x^4 e^x}{(e^x-1)^2} dx} \quad (4)$$

where $x = \hbar\omega/k_B T$ the normalized phonon frequency, $\omega$ the phonon frequency, $\theta_i$ the Debye temperature of the *i*-th phonon branch. $\tau_{N,i}$ is the relaxation time for Normal 3-phonon process, $\tau_{R,i}$ is the relaxation time for resistive scattering process, and $\tau_{C,i}^{-1} = (\tau_{N,i}^{-1} + \tau_{R,i}^{-1})$. $\tau_{R,i}$ is contributed by several different resistive processes, such as phonon-phonon umklapp scattering ($\tau_{U,i}$), point scattering ($\tau_{P,i}$) and boundaries scattering ($\tau_{B,i}$), and it can be expressed using Matthiessen's rule:

$$\tau_{R,i}^{-1} = (\tau_{U,i}^{-1} + \tau_{P,i}^{-1} + \tau_{B,i}^{-1}) \tag{5}$$

Each phonon scattering process is further discussed here. $\tau_{N,i}$ are given by Asen-Palmer [29]:

$$\tau_{N,L} = 1 / BNL * \omega^2 * T^3 \tag{6}$$

$$\tau_{N,T} = 1 / BNT * \omega * T^4 \tag{7}$$

where $BNL = \frac{K_B^3 \gamma_L^2 V}{M \hbar^2 v_L^5}$, $BNT = \frac{K_B^4 \gamma_T^2 V}{M \hbar^3 v_T^5}$, M is the atomic mass and V is the atomic volume; $\gamma_L$ and $\gamma_T$ are the Grüneisen parameters for the longitudinal and transverse branches.

$\tau_{U,i}$ are given as [30]:

$$\tau_{U,L} = 1 / BULa * \omega^2 * T * e^{-\theta_L/3T} \tag{8}$$

$$\tau_{U,T} = 1 / BUTa * \omega^2 * T * e^{-\theta_T/3T} \tag{9}$$

where $BULa = \frac{\gamma_L^2 \hbar}{v_L^2 M \theta_L}$, $BUTa = \frac{\gamma_T^2 \hbar}{v_T^2 M \theta_T}$.

$\tau_{P,i}$ is given with the Klemens formalism as [47]:

$$\tau_{P,L} = (4 * \pi * v_L^3) / (V * \Gamma * \omega^4) \tag{10}$$

$$\tau_{P,T} = (4 * \pi * v_T^3) / (V * \Gamma * \omega^4) \tag{11}$$

where $\Gamma$ is the point scattering factor, which is discussed later in this paper.

$\tau_{B,i}$ is given as [48]:

$$\tau_{B,L} = d / (v_L) \tag{12}$$

$$\tau_{B,T} = d / (v_T) \tag{13}$$

where *d* is the effective length of the boundary.

Note that the model used here is similar to the models [31-33] used for $Al_{1-x}Ga_xN$ and $Si_xGe_{1-x}$. These models assume a dispersionless Debye system where the low frequency (long wavelength) phonons contribute more significantly to the thermal transport [32]. Based on the model described, predicting the thermal conductivity of $Al_{1-x}Sc_xN$ (*k*) needs the parameters of the alloy including: sound velocity ($v_L$ and $v_T$), atomic mass (*M*), atomic length (*δ*), Grüneisen parameters ($\gamma_L$ and $\gamma_T$), Debye temperatures ($\theta_L$ and $\theta_T$), point scattering factor ($\Gamma$) and effective boundary length (*d*). Among them, $v_L, v_T$, M, $\delta$, $\gamma_L$, $\gamma_T$, $\theta_L$ and $\theta_T$ are a function of x and obtained by the virtual crystal model [31, 32], with the calculation details seen in the **supplementary material.** *d* is assumed to be the lateral grain size (*i.e.,* 100 nm) for polycrystalline $Al_{1-x}Sc_xN$. This assumption has been successfully applied for various polycrystalline dielectrics [42, 49, 50]. $\Gamma$ is affected by many factors, such as different isotopes ($\Gamma_{iso}$), impurities/vacancies ($\Gamma_{im}$), and alloy ($\Gamma_{alloy}$). We have found that $\Gamma_{iso}$ and $\Gamma_{im}$ contribute negligibly to the reduction in $k$ for $Al_{1-x}Sc_xN$ (see the details in **supplementary material**). Thus, $\Gamma$ is assumed to be only determined by $\Gamma_{alloy}$ which can be expressed as [31, 32]:

$$\Gamma_{alloy} = \sum_j x_j \left\{ [(M_j - M)/M]^2 + \varepsilon[(\delta_j - \delta)/\delta]^2 \right\} \tag{14}$$

where $j$ is the index for components ScN or AlN. The first term in the brackets is the mass-difference-induced scattering factor and the second term is the one caused by the strain field difference. $\varepsilon$ is a phenomenological parameter and has been estimated to be 39 for $Si_xGe_{1-x}$ [31] and $Al_{1-x}Ga_xN$ [32]. However, this value is not applicable for $Al_{1-x}Sc_xN$. As shown in Figure 2 (a), the modelled $k$ with $\varepsilon$ being 39 overestimate all measured results. Therefore, in our study $\varepsilon$ is treated as an adjustable parameter that is used to obtain the fit between the model prediction and the experimental data. As shown in Figure 2(a), the best fit result of $\varepsilon$ is 200.

After determining $\varepsilon$, the thermal conductivity ($k$) modelled as function of temperature is plotted against the experimental results in Figures 2(b). Good quantitative agreement between theory and experiment is obtained, with the exception of $Al_{1-x}Sc_xN$ (x=0.028 and 0.076) samples at low temperatures (80K-170K). At 80K-170K range, heat penetration depth during TDTR characterization (defined as $h \sim \sqrt{k/\pi f C}$, where $f$ is the TDTR pump beam modulation frequency and $C$ is the volumetric heat capacity) is about 400nm-700nm for $Al_{1-x}Sc_xN$ (x=0.028 and 0.076) samples, exceeding the samples thickness (400 nm). At higher temperature (>170 K), $h$ is well below the thickness. Thus, discrepancies at the low temperatures are likely due to the defects (smaller grains, minimal twists and orientation disturbance) observed near the $Al_{1-x}Sc_xN$/Si interface not included in theoretical calculations. In Figure 2(b), we further plotted $k$ as function of temperature with the $d$ of 75nm, assuming that near interface defects have an influence on reducing $d$. The modified modelling results do match better with experiments in the lower temperature range, supporting our assumption for the observed discrepancies.

To understand the effects of different scattering processes on $k$, we analyze the spectral thermal conductivity and phonon relaxation time. The inset of Figure 3(a) plots the spectral contribution to thermal conductivity for a $Al_{1-x}Sc_xN$ (x=0.173) film, with the calculation method provided in [33]. The spectral curve increases with frequency reaching a peak at around 20 THz and decreases thereafter. This demonstrates that low frequency (<40THz) phonons contribute more significantly to the transport and thus the treatment of alloys as a dispersionless Debye-like system is valid. This is similar to what was found in $Si_xGe_{1-x}$ [31]. The normal-3 ($\tau_{N,i}$), umklapp ($\tau_{U,i}$), alloy ($\tau_{A,i}$) and boundary ($\tau_{B,i}$) relaxation times of longitudinal and transversal models are plotted in Figure 3 (a) and 3(b), respectively. Also plotted is the total relaxation time $\tau_{C,i}^{-1}$, which is expressed using Matthiessen's rule:

$$\tau_{C,i}^{-1} = \left( \tau_{N,i}^{-1} + \tau_{U,i}^{-1} + \tau_{A,i}^{-1} + \tau_{B,i}^{-1} \right) \tag{15}$$

The results reveal that in the frequency range below approximately 20 THz, the total relaxation time is limited by $\tau_{B,i}$; above that frequency, $\tau_{A,i}$ becomes the dominate limiting factor. Therefore, the low thermal conductivities of $Al_{1-x}Sc_xN$ thin films are primarily due to boundary and alloy scattering. This interpretation can be further demonstrated through the temperature dependence of the thermal conductivity presented in Figure 2(b). Over the whole temperature region, the thermal conductivity of bulk single crystal AlN show a descent trend indicative of strong normal-3 and umklapp scatterings. The temperature-dependent behavior in the $Al_{1-x}Sc_xN$ is obviously different owing to the strong boundary and alloy scattering.

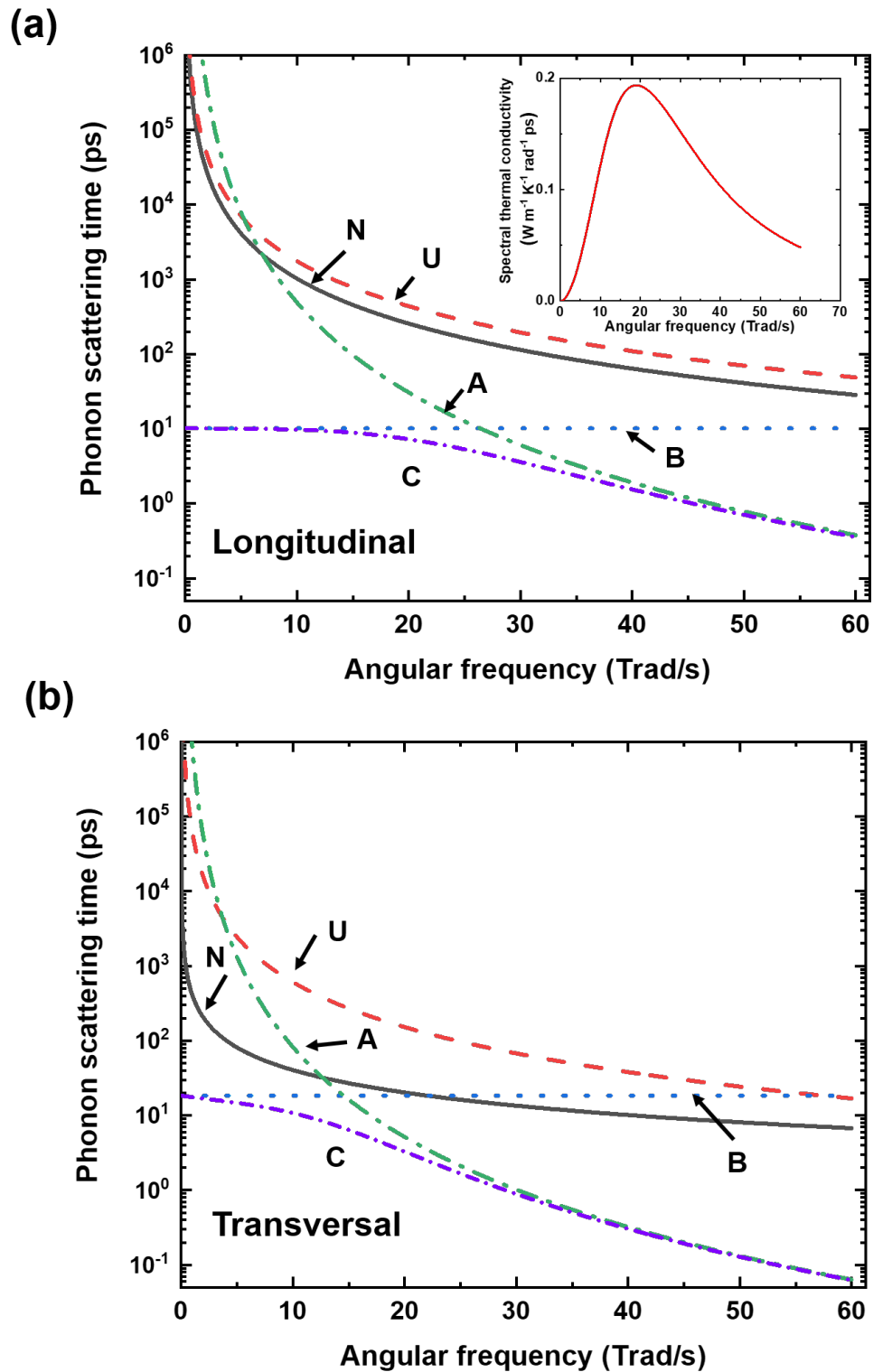

Figure 3. The phonon relaxation times of (a) longitudinal and (b) transversal models versus angular frequency. N, U, A, B and C indicates the normal-3, umklapp, alloy, boundary and total

relaxation times, respectively. The inset in (a) shows the spectral thermal conductivity at room temperature. Results in (a) and (b) are calculated from a Al$_{1-x}$Sc$_x$N (x=0.173) film.

Next, we focused on the effect of alloy composition, x. Based on the virtual crystal model (see details in Eq. (S3) to (S6) in the **supplementary material)**, sound velocities of Al$_{1-x}$Sc$_x$N ($v_L$ and $v_T$) reduce with the x increasing. Since *k* is proportional to sound velocities as seen in Eq.(1), *k* reduction with increasing x is partially due to the decrease of $v_L$ and $v_T$. Another primary impact of x increasing is the increase of alloying scattering, further leading to the reduction in *k*. This can be validated with Figure 4, which shows an increased alloying scattering factor, $\Gamma_{alloy}$, as a function of x. Figure 4 also compares the factor contributed by mass-difference ($\Gamma_M$) and strain-field-difference ($\Gamma_S$). It is seen that $\Gamma_S$ is significantly larger than $\Gamma_M$, and dominates the total alloy scattering factor $\Gamma_{alloy}$. This is in contrast to the Al$_{1-x}$Ga$_x$N and Si$_x$Ge$_{1-x}$ alloy systems where $\Gamma_M$ dominates (see the discussion details in the **supplementary material**). Moreover, we found that in the Al$_{1-x}$Ga$_x$N and Si$_x$Ge$_{1-x}$ systems, thermal conductivity prediction without including $\Gamma_S$ show negligible difference (see the calculation details in **supplementary material**). However, it is not the case for Al$_{1-x}$Sc$_x$N studied here, demonstrating that the alloy scattering processes in Al$_{1-x}$Sc$_x$N is significantly different to others.

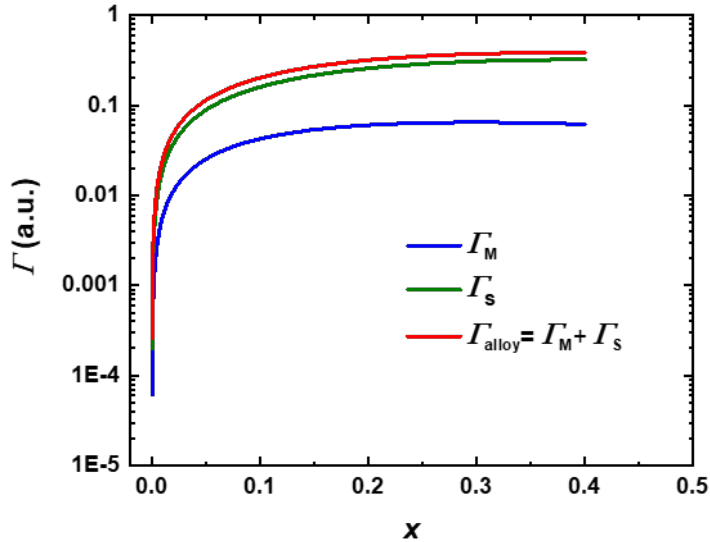

Figure 4. Alloy scattering factor, $\Gamma_{alloy}$, as a function of Sc composition *x*, and its contributions from mass-difference ($\Gamma_M$) and strain-field-difference ($\Gamma_S$).

With the phonon/thermal properties data (*e.g.* thermal conductivity, phonon relaxation time, Grüneisen parameters, and phonon group velocities), one can predict the Akhiezer damping limits from Boltzmann transport equations [6, 8, 51], as well as the Landau-Rumer and thermoelastic damping limits [7, 52, 53]. This has generally been applied for various resonators made by single crystal materials (assuming perfect crystal conditions), such as AlN, Si, SiC, diamond, and GaN. [7, 8, 52, 53], to predict the intrinsic *Q* limit, determined by the intrinsic phonon-phonon scattering solely. This, however, has not been applied for resonators made of Al$_{1-x}$Sc$_x$N in which there are additional scattering processes, *i.e.*, phonon-alloy scattering naturally present due to the inherent crystal structure, and phonon-boundary scattering due to the columnar grain structures formed in the crystals. Phonon interactions with alloys and boundary defects ultimately affect the *Q*, and therefore require precise characterization. The explicit analysis of various scattering processes here, as well as the measured thermal conductivities, shed light on the further *Q* studies in $Al_{1-x}Sc_xN$ acoustic devices. In addition, the measured thermal conductivities enable the quantitative analysis of heat transfer and thermal-mechanical processes in resonator devices operating under high power loads, further benefitting the devices design.

We report the thermal conductivity and dominant phonon scattering processes in epitaxial $Al_{1-x}Sc_xN$ alloy thin films. Thermal conductivities of the films, measured by time-domain thermoreflectance at room temperature, were found to decrease with increasing alloy composition in the range of x from 0.028 to 0.26. Temperature-dependent measurements showed that thermal conductivity increases with temperature from 80 K to 200 K, and then keep constant for the higher temperatures up to 480 K. A virtual crystal analytical model well reproduced the measured x- and temperature-dependent thermal conductivities. Application of the model enables us to elucidate the boundary and alloy scattering effects on limiting the thermal conductivities. We also highlighted that the alloy scattering in $Al_{1-x}Sc_xN$ is caused mainly by strain-field-difference, and not as much by the atomic-mass-difference between ScN and AlN. This work provides quantitative information for thermal conductivities and useful insights for phonon scatterings mechanisms in $Al_{1-x}Sc_xN$, open perspectives in understanding and controlling the energy dissipation and performance in acoustic resonators.

See supplementary material for the details of X-ray photoelectron spectroscopy (XPS) experiments, time-domain thermoreflectance (TDTR) experiments, virtual crystal model, isotopes scattering ($\Gamma_{iso}$) and impurities scattering ($\Gamma_{im}$) factors, and $Al_{1-x}Ga_xN$ and $Si_xGe_{1-x}$ thermal conductivity calculations.

The data that support the findings of this study are available from the corresponding author upon reasonable request.

Acknowledgement
M.P., Y.Z., and A.A. acknowledge financial support from IQE plc.

# Supplemental Material

# Phonon heat conduction in $Al_{1-x}Sc_xN$ thin films

Chao Yuan[1], Mingyo Park[2], Yue Zheng[2], Jingjing Shi[1], Rytis Dargis[3], Samuel Graham[1]*, Azadeh Ansari[2]*

1 George W. Woodruff School of Mechanical Engineering, Georgia Institute of Technology, Atlanta, Georgia 30332, USA
2 School of Electrical and Computer Engineering, Georgia Institute of Technology, Atlanta, Georgia 30332, USA
3 IQE NC, 494 Gallimore Dairy Rd., Greensboro, NC, 27407 USA

*Author to whom correspondence should be addressed: sgraham@gatech.edu; azadeh.ansari@ece.gatech.edu

## S1. X-ray photoelectron spectroscopy (XPS) experiments

The Sc concentration (x) of the nitride layers was confirmed by X-ray photoelectron spectroscopy (XPS) with high-resolution scans using a thermo K-Alpha XPS. Samples surfaces were sputtered clean prior to XPS characterization. x in $Al_{1-x}Sc_xN$ is evaluated by the formula $\frac{\text{Sc molar faction}}{\text{Sc molar faction + Al molar faction}}$ based on the molar fractions of the Sc, Al and N detected by XPS. In addition to compositions of Sc, Al and N, composition of O was detected near the surface region with <8% O concentration ($\frac{\text{O molar fraction}}{\text{total}}$). Such O detection cannot be avoided due to the natural surface oxidation by the residual O from the sputter gun and the measurement chamber. Thus, the relative high O concentration discovered near surface region does not represent the genuine O concentration for the as-grown $Al_{1-x}Sc_xN$ films.

## S2. Time-domain thermoreflectance (TDTR) experiments

We used a two-color pump-probe TDTR setup, where a modulated 400 nm pump beam periodically heats the metal transducer, while a delayed 800 nm probe beam monitors transducer surface reflectance change induced by the surface temperature change. More configuration details of our TDTR system can be found elsewhere [1, 2]. For TDTR transduction, the surfaces of $Al_{1-x}Sc_xN$ on Si samples were coated with Au(75nm)/Ti(5nm) via e-beam evaporation. We measured the TDTR signal (the $-V_{in}/V_{out}$ ratio in this study) from the Au/Ti coated samples, with a typical measured signal shown in Figure S1(a). This signal contains the thermal properties and structural information of the Au/Ti- $Al_{1-x}Sc_xN$-Si stack, and the pump and probe beams characteristics. Among them, temperature-dependent thermal conductivities and volumetric heat capacities of metal transducer, temperature-dependent volumetric heat capacities of $Al_{1-x}Sc_xN$ and Si, the thicknesses of each layer, pump beam modulating frequency, and spot sizes of two beams are given as the known parameters. The remaining unknown parameters, the thermal boundary conductance at Au/Ti-$Al_{1-x}Sc_xN$ interface ($TBC_{Au/Ti}$), $Al_{1-x}Sc_xN$ thermal conductivity ($k_{AlScN}$), TBC at $Al_{1-x}Sc_xN$-Si interface (TBC$_{AlScN}$) and Si thermal conductivity ($k_{Si}$), are treated as the free variables to fit the analytical model to the measured signal.

Figure S1 (a) and S1 (b) present the sensitivity plots for the Au/Ti- Al$_{1-x}$Sc$_x$N -Si (x=7.6%) sample at 292 K and 80 K, respectively, with the pump beam modulating frequency being 8.8 MHz. At 292 K, the measurement sensitivities to TBC$_{AlScN}$ and $k_{Si}$ are negligible since the heat penetration depth $h \sim \sqrt{k/\pi f C}$ is approximately 330 nm which is below the Al$_{1-x}$Sc$_x$N thickness. In this calculation, $k$ is the through-plane thermal conductivity, $f$ is the modulation frequency and $C$ is the volumetric heat capacity. Thus, TBC$_{AlScN}$ and K$_{Si}$ can be safely ignored, and TBC$_{Au/Ti}$ and $k_{AlScN}$ are the remained unknown parameters and obtained by the fitting, as seen in Figure S1 (a). In fact, from 210K and above, the sensitivity plots for all Au/Ti- Al$_{1-x}$Sc$_x$N -Si samples are similar to that shown in Figure S1 (b). Thus, data analysis of them is as described above. However, when the temperature is below 210K, for example at 80K, the signal is sensitive to all four parameters, as seen in Figure 2(c). This is because the penetration depth ($h \sim \sqrt{k/\pi f C}$~750 nm) exceeds the thickness. Thus, all the four parameters are treated as free adjustable parameters in data fitting. It should be noted that the sensitivities to TBC$_{AlScN}$ and K$_{Si}$ are still small, in contrast to the sensitivity to k$_{AlScN}$. Accordingly, the fitted results of TBC$_{AlScN}$ and K$_{Si}$ are usually some kind of arbitrary values. However, this doesn't affect the reliable fitting for k$_{AlScN}$ as it dominates the measurement sensitivity across the time regime. This has also been verified by the truth that setting the TBC$_{AlScN}$ and K$_{Si}$ with different random values in the data fitting, the fitted k$_{AlScN}$ value barely changes. Note that although we measured the TDTR signal from sub-picsecond to 6 nanoseconds, the starting point of the data for fitting was chosen as 300 ps to avoid any electron-phonon coupling effects [3].

At least three measurements were taken on each sample at different locations to ensure relative uniformity. The fitted TBC$_{Au/Ti}$ ranges about 30-70 MW/m$^2$K for different samples and temperatures. The obtained $k_{AlScN}$ results has been plotted in Figure 2 (a) and 2 (b) in the main body of the paper. The error was obtained by a Monte Carlo technique [1], accounting for the uncertainties in Al$_{1-x}$Sc$_x$N film thickness (+/-5%) and volumetric heat capacities (+/-2%), in Au/Ti thickness (+/-5%), thermal conductivities (+/-10%) and volumetric heat capacities (+/-2%), and in pump/probe beam sizes (+/-3%). Based upon a 95th/5th percentile confidence interval, it reports a ~ +/-18% error bar for $k_{AlScN}$ at 292 K, and a ~ +/-27% error bar at 80K.

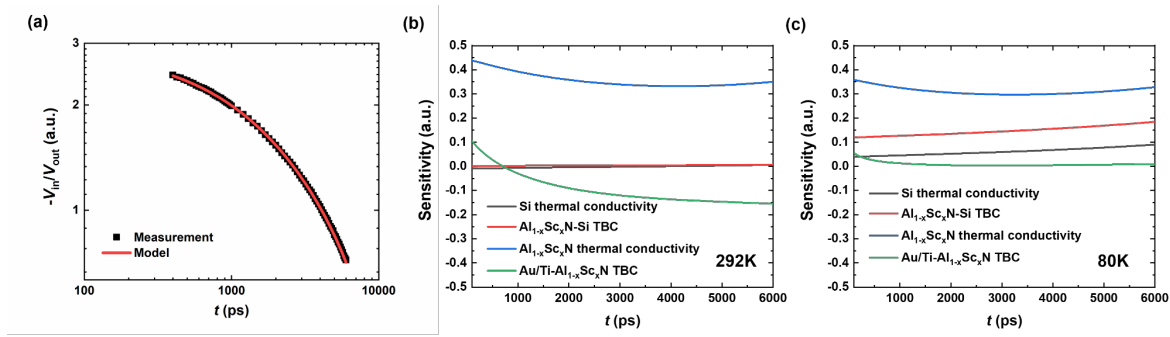

Figure S1. (a) TDTR trace (the -$V_{in}$/$V_{out}$ ratio in this study) measured and modelled for a Au/Ti coated Al$_{1-x}$Sc$_x$N (x=0.076) sample. Measurement sensitivity plots for the Au/Ti- Al$_{1-x}$Sc$_x$N -Si (x=0.076) sample at (b)292 K and (c)80 K, respectively.

## S3. Virtual crystal model

The basic idea of the virtual crystal model is to replace the disordered lattice by an ordered virtual crystal with randomly distributed atoms of constituent materials. The average lattice parameters and phonon-related properties, such as sound velocity ($v_L$ and $v_T$), atomic mass (M), atomic volume (V), Grüneisen parameters ($\gamma_L$ and $\gamma_T$), and Debye temperatures ($\theta_L$ and $\theta_T$), are taken according to the volumetric composition of the material.

The virtual atomic weight M is assumed to be the mass average of different components of the alloy:

$$M = mM_{ScN} + (1-m)M_{AlN} \quad (S1)$$

where m is the mass fraction of ScN component in the alloy. The virtual atomic length is given by Vegard's law through the equation

$$\delta = m\delta_{ScN} + (1-m)\delta_{AlN} \quad (S2)$$

where $\delta = V^{1/3}$ and V is the atomic volume.

Based on the M and V of AlN and ScN, given in Table S1, M, V and $\delta$ of Al$_{1-x}$Sc$_x$N can be obtained by equations (S1)-(S2). Noticed that for the x within the studied range (0<x<0.26), the $\delta$ calculated based on virtual crystal model is actually consistent with the density functional theory (DFT) calculation based on a more realistic lattice structure created [4].

The expressions for longitudinal $v_L$ and transverse ($v_{T1}, v_{T2}$) sound velocities can be written as

$$v_L = \sqrt{c_{11}/\rho_{AlScN}} \quad (S3)$$
$$v_{T1} = \sqrt{c_{44}/\rho_{AlScN}} \quad (S4)$$
$$v_{T2} = \sqrt{(c_{11}-c_{12})/2\rho_{AlScN}} \quad (S5)$$

where $c_{11}$, $c_{12}$ and $c_{44}$ are elastic constants, $\rho_{AlScN}$ is the density of Al$_{1-x}$Sc$_x$N. These formulas (S3-S5) are generally assumed for the wurtzite structure material [5]. While Al$_{1-x}$Sc$_x$N consists of the rocksalt ScN and wurtzite AlN, the Al$_{1-x}$Sc$_x$N materials with x less than 0.4 was found to exhibit a single-phase wurtzite structure. Therefore, the formulas (S3-S5) are applicable for the Al$_{1-x}$Sc$_x$N studied here. Keyes [6] proposed a formula used to obtained the elastic constants $c_{ik}$:

$$c_{ik}\delta^4 \approx const \quad (S6)$$

From this formula, and from the known elastic constants and atomic volume of AlN, we can obtain the elastic constants for the virtual crystal. The elastic constants $c_{ik}$ have also been calculated by using DFT [7]. We find that the DFT results and results by the Keyes approach match well with the discrepancy less than 3% for x<0.26.

From Ref. [5], the Debye temperature ($\theta$) can be expressed as:

$$\theta \propto (\delta c_{ik}/M)^{1/2} \quad (S7)$$

Combine with (S6), we obtain

$$\theta M^{1/2}\delta^{3/2} = \alpha \quad (S8)$$

We followed the assumption made in [5] that $\alpha$ is a constant. $\alpha$ is first determined from the $\theta$, M and $\delta$ of AlN. This $\alpha$ value is then used to calculate the $\theta$ of the virtual crystal. Following Ref. [8], the AlN longitudinal and transverse Debye temperatures ($\theta_L$ and $\theta_T$) were determined by:

$$\theta_L = \frac{\hbar F_L}{K_B} \quad \text{(S9)}$$

$$\theta_T = \frac{\hbar F_T}{K_B} \quad \text{(S10)}$$

where $F_L$ and $F_T$ are the zone-boundary frequencies of L and T branches, and equal to be 10.5 and 5.3 THz [9], respectively.

In our model, we assume $Al_{1-x}Sc_xN$ keeps the same Grüneisen parameters ($\gamma_L$ and $\gamma_T$) of AlN. $\gamma_L$ and $\gamma_T$ of AlN were obtained by model fit with single crystal AlN thermal conductivities results [9]. The fitted $\gamma_L$ and $\gamma_T$ results are summarized in Table S1.

Table S1 Materials properties for AlN, ScN, GaN, Si and Ge

| Material | V (m³/atom) | ρ (kg/m³) | M (kg/atom) | $C_{11}$ (m/s) | $C_{12}$ (m/s) | $C_{44}$ (m/s) | $F_L$ (THz) | $F_T$ (THz) | $\theta_L$ (K) | $\theta_T$ (K) | $\gamma_L$ | $\gamma_T$ |
|---|---|---|---|---|---|---|---|---|---|---|---|---|
| AlN | 1.05 $10^{-29}$ | 3266 | 3.42 $10^{-26}$ | 397 [7] | 137 [7] | 118 [7] | 10.5 [9] | 5.3 [9] | 506 | 252 | 0.74 | 0.43 |
| ScN | 1.33 $10^{-29}$ | 4250 | 5.67 $10^{-26}$ | - | - | - | - | - | - | - | - | - |
| GaN | 1.12 $10^{-29}$ | 6150 | 6.91 $10^{-26}$ | - | - | - | - | - | - | - | - | - |
| Si | 2 $10^{-29}$ | 2200 | 4.40 $10^{-26}$ | 166 [10] | 64 [10] | 80 [10] | 12.4 [8] | 5.1 [8] | 586 | 240 | 1.1 [8] | 0.6 [8] |
| Ge | 2.26 $10^{-29}$ | 5323 | 1.20 $10^{-25}$ | - | - | - | - | | - | | - | |

### S4. $\Gamma_{iso}$ and $\Gamma_{im}$

The phonon-isotope scattering factor for N element ($\Gamma_{iso-N}$) made up of two naturally occurring isotopes (99.63% N14, 0.37% N15) was $1.9*10^{-5}$ [8]. Naturally occurring scandium is composed of only one stable isotope, Sc45, so does naturally occurring Al (only Al27 isotope is stable). The average phonon-isotope scattering factor ($\Gamma_{iso}$) of the compound $Al_{1-x}Sc_xN$ is given by the stoichiometry weighted average of each element's phonon-isotope factor [8]. Therefore, $Al_{1-x}Sc_xN$ has the $\Gamma_{iso}$ less than $1.9*10^{-5}$.

$\Gamma_{im}$ can be calculated based on fractional concentrations of impurities and vacancies atoms inside the materials [8]. Although we detected the O impurity near surface region by XPS characterization, we have concluded that the detected impurities do not represent the genuine O concentration for the as-grown $Al_{1-x}Sc_xN$ films (See details in S1). Thus, we refer to the literatures reported values. We noted that $\Gamma_{im}$ of a magnitude of $10^{-4}$ have been found for the III-V compounds (e.g. GaN and AlN) [5]. In this study, we assumed a $\Gamma_{im}$ of $3*10^{-4}$ [5] for the $Al_{1-x}Sc_xN$ samples.

With Eq. (14) given in the main body of the paper, the alloy scattering factor $\Gamma_{alloy}$ ranges from 0.07 to 0.35 for x from 0.028 to 0.2. Thus, $\Gamma_{alloy}$ is 2-3 order of magnitude larger than $\Gamma_{im}$, and 4-5 order of magnitude larger than $\Gamma_{iso}$. Figure S2 further shows that the isotopes and vacancies/impurities effects contribute negligibly to the reduction in thermal conductivity.

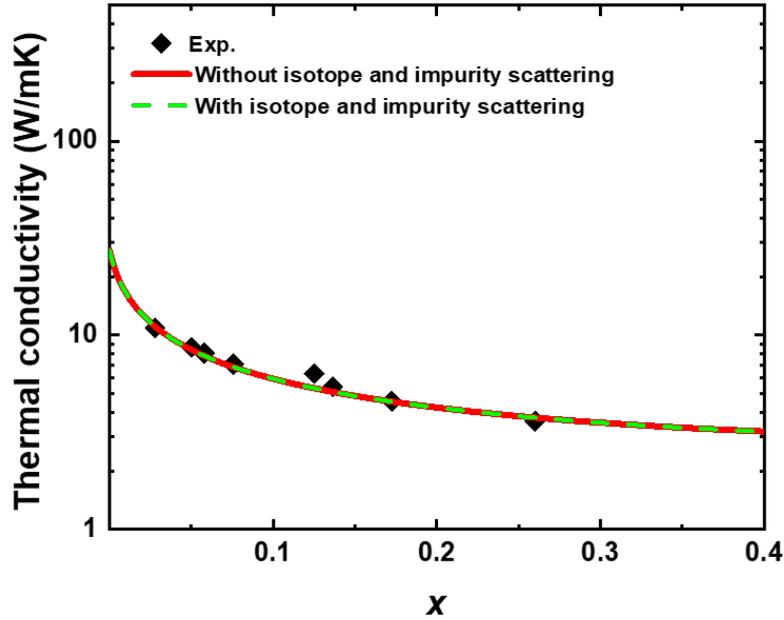

Figure S2. Calculated thermal conductivities by including and excluding isotope and impurities scattering, showing that the isotopes and impurities scatterings contribute negligibly to the reduction in thermal conductivity.

**S5. $Si_xGe_{1-x}$ and $Al_{1-x}Ga_xN$ thermal conductivity calculations**

$Si_xGe_{1-x}$ and $Al_{1-x}Ga_xN$ thermal conductivities were calculated with the same way for $Al_{1-x}Sc_xN$, using the modified Callaway-Holland model with the inputs obtained by the virtual crystal method. The parameters used for $Si_xGe_{1-x}$ and $Al_{1-x}Ga_xN$ thermal conductivities calculations are summarized in table S1. Phenomenological strain parameter ($\varepsilon$) of 39 [5, 11] was applied for both $Si_xGe_{1-x}$ and $Al_{1-x}Ga_xN$ calculations. Figures S3(a) and S3(b) plot the calculated thermal conductivities for $Si_xGe_{1-x}$ and $Al_{1-x}Ga_xN$, respectively, along with literature experimental values. The boundary lengths in model for $Si_xGe_{1-x}$ and $Al_{1-x}Ga_xN$ are given as 200 nm and 400 nm respectively, corresponding to average boundary lengths of experimental samples. The model results match quantitatively well with experimental results, further validate the virtual crystal based Callaway-Holland model.

With the model, we calculated alloying scattering factor, $\Gamma_{alloy}$, as a function of x, and its contributions from mass difference ($\Gamma_M$) and strain field difference ($\Gamma_S$) for $Si_xGe_{1-x}$ and $Al_{1-x}Ga_xN$. The results are plotted in Figure S3 (c) and S3(d), respectively, demonstrating that $\Gamma_S$ is significantly lower than $\Gamma_M$ in both $Si_xGe_{1-x}$ and $Al_{1-x}Ga_xN$ systems. This is in contrast to $Al_{1-x}Sc_xN$ system that $\Gamma_S$ is significantly larger than $\Gamma_M$ (See the details in the main body of the paper). In Figure S3 (a) and S3 (b), thermal conductivities of $Si_xGe_{1-x}$ and $Al_{1-x}Ga_xN$ modeling by excluding $\Gamma_S$ were also plotted, further illustrating that $\Gamma_S$ has negligible effect in $Si_xGe_{1-x}$ and $Al_{1-x}Ga_xN$.

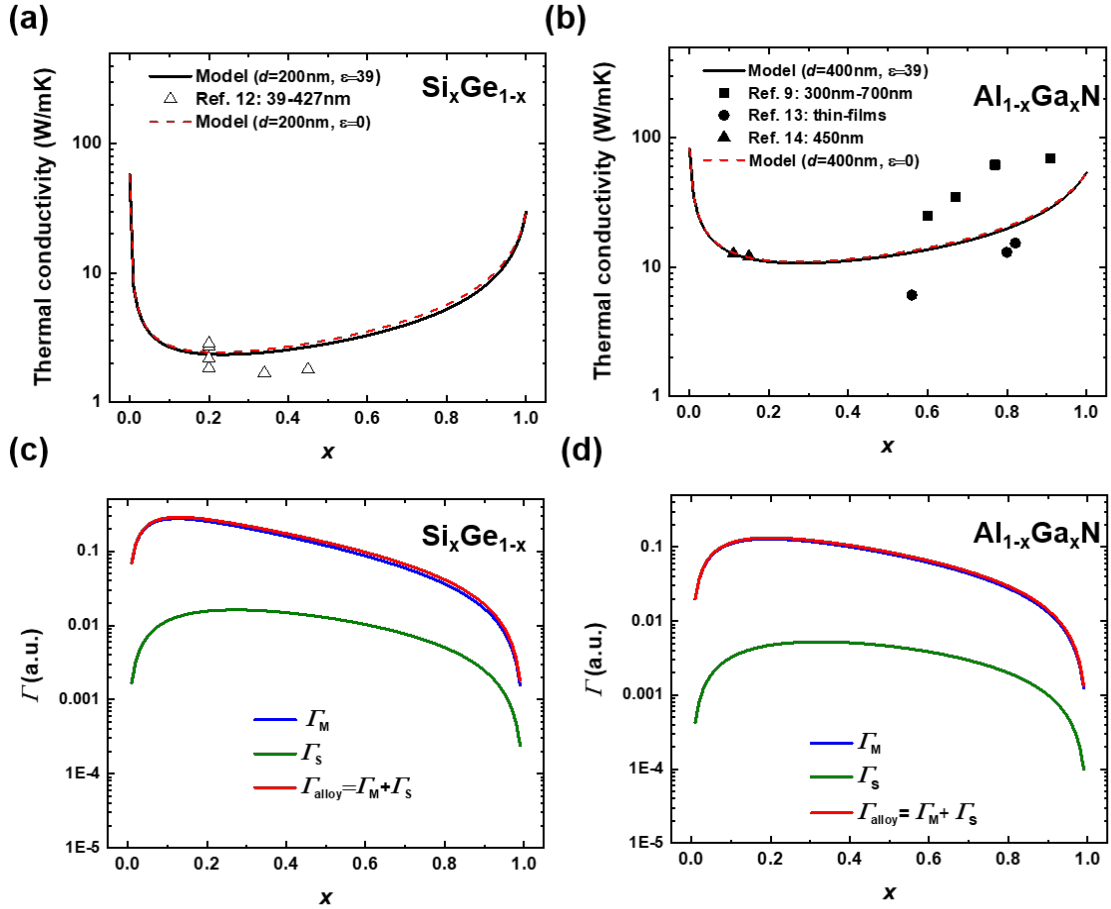

Figure S3. Thermal conductivities of (a) $Si_xGe_{1-x}$ and (b) $Al_{1-x}Ga_xN$ calculated as function of $x$, along with literature experimental values [5, 12-14] for comparison. Alloy scattering factor, $\Gamma_{alloy}$, as a function of $x$, and its contributions from mass difference $\Gamma_M$ and strain field difference $\Gamma_S$ for(c) $Si_xGe_{1-x}$ and (d) $Al_{1-x}Ga_xN$.